**Moving towards FAIR practices in epidemiological research**

Montserrat García-Closas[1], Thomas U. Ahearn[1], Mia M. Gaudet[1], Amber N. Hurson[1], Jeya Balaji Balasubramanian[1], Parichoy Pal Choudhury[1], Nicole M. Gerlanc[1], Bhaumik Patel[1], Daniel Russ[1], Mustapha Abubakar[1], Neal D. Freedman[1], Wendy S.W. Wong[1], Stephen J. Chanock[1], Amy Berrington de Gonzalez[1*], Jonas S Almeida[1*]

[1] Division of Cancer Epidemiology and Genetics, National Cancer Institute, USA

*Contributed equally

**Corresponding author:**

Montserrat Garcia-Closas, MD, MPH, DrPH
Trans-Divisional Research Program
Division of Cancer Epidemiology and Genetics
National Cancer Institute
9609 Medical Center Drive
Rockville, MD 20850
USA

Email: montserrat.garcia-closas@nih.gov

**Abstract**

Reproducibility and replicability of research findings are central to the scientific integrity of epidemiology. In addition, many research questions require combiningdata from multiple sources to achieve adequate statistical power. However, barriers related to confidentiality, costs, and incentives often limit the extent and speed of sharing resources, both data and code. Epidemiological practices that follow FAIR principles can address these barriers by making resources (F)indable with the necessary metadata , (A)ccessible to authorized users and (I)nteroperable with other data, to optimize the (R)e-use of resources with appropriate credit to its creators. We provide an overview of these principles and describe approaches for implementation in epidemiology. Increasing degrees of FAIRness can be achieved by moving data and code from on-site locations to the Cloud, using machine-readable and non-proprietary files, and developing open-source code. Adoption of these practices will improve daily work and collaborative analyses, and facilitate compliance with data sharing policies from funders and scientific journals. Achieving a high degree of FAIRness will require funding, training, organizational support, recognition, and incentives for sharing resources. But these costs are amply outweighed by the benefits of making research more reproducible, impactful, and equitable by facilitating the re-use of precious research resources by the scientific community.

**Introduction**

Reproducibility and replicability of research findings is central to scientific integrity; how to improve it in epidemiology has been debated for decades[1,2]. Over the last 15 years, epidemiology has seen a growing number of consortia of studies to improve replication and attain large sample sizes. These efforts require sharing of research resources, both data and code. Yet, there are factors that limit the extent and speed of data sharing, including (1) data confidentiality concerns, (2) effort and cost required to make complex data usable by others, and (3) minimal incentives for data creators to share resources, owing to concerns about competitionfrom other groups and inadequate credit to the resource creators who invested significant energy and funds generating the data[3-6]. Large scale, high dimensional data currently beinggnerated in epidemiology (e.g., by "omic" technologies, wearable devices, imaging and electronic health records) provide great opportunities for improving exposure and outcome assessment; however, they create even more daunting challenges for data management, storing, and sharing.

Epidemiological practices that follow FAIR principles[7] can address these challenges by making research resources (F)indable online with the necessary metadata , (A)ccessible to authorized users and (I)nteroperable with other resources, to optimize the (R)e-use of the scientific resources with appropriate credit to its creators. While FAIR principles are often applied to data, they also apply to analytical code or software[8]. The need for rapid generation and sharing of data and research tools to fight COVID-19 has recently accelerated the development of FAIR data systems[9,10]. FAIR principles have been rapidly and broadly adopted of by many funding agencies, including the US National Institutes of Health (NIH)[11,12] and the European Commission (EC)[13]. The requirements for submission of data management and sharing plans following FAIR principles are already in effect in the EU and coming into effect in 2023 for the NIH[12]. In addition, an increasing number of scientific journals are establishing data policies to encourage data and code sharing[14]. Consequently, adoption of FAIR principles has become a high priority.

Drawing on an evolving area of research in data science [15,16], this article provides an overview of FAIR principles in the context of epidemiological research. We explain how increasing compliance with these principles can improve the quality and reproducibility of research and facilitate data access and sharing. We describe approaches for implementation of FAIR principles and illustrate them with examples of current and emerging resources. In conjunction with rigorous methodology, such systems promise to accelerate all areas of epidemiological research.

**What do FAIR principles mean in epidemiology?**

This section describes FAIR principles when applied to data and the code to clean, manage and analyze it (**Figure 1**), and provides a non-technical description of what these principles mean for epidemiological research (see **Box 1** for definitions of terms).

*Findable*

Researchers need to know the location of research resources to find them along with rich descriptors to enable their appropriate use. This is best achieved when resources (i.e., metadata, data and code) can be found remotely online from any location. Importantly, the concept of findability applies both to resources that can be found by anyone through a Web search and to resources that can only be found by those with permissions (e.g., it applies to both open and restricted access datasets).

Moving resources from on-site servers or computers to the Cloud greatly facilitates findability as it enables locating resources online using Web addresses (i.e., Uniform Resource Locators (URLs)). Ideally, URLs should be "globally unique and persistent identifiers" (PIDs), which are a type of URL that is permanent and uniquely identifies the location of the resources on the Web. This type of URL never changes, ensuring links do not break, even if websites are updated. Examples of PIDs include Digital Object Identifiers (DOIs) to uniquely find and cite publications or researchers (e.g., ORCID iD). Similarly, PIDs can be used to uniquely identify datasets, metadata or code, so that they can be easily found and cited. PIDs are assigned when resources are moved to the Cloud, either by using consumer-facing data storage and management services (e.g., URL link automatically generated when moving a file to Google Drive), or assigned by searchable online data or code repositories such as GitHub. Repositories that assign PIDs to datasets rather than "locally-unique" accession numbers (e.g., those created by dbGap) are thus preferable to make resources findable. Guidance on how to cite data and link to publications is provided by organizations such as the FORCE11 Data Citation Synthesis Group[17,18].

*Accessible*
Once researchers locate research resources, they should know how to access them securely over the internet according to all applicable permissions. This is not the same as open access, rather there should be a clear process for access according to the appropriate licenses, data use agreements (DUAs), or other data governance requirements. For datasets, this includes information on who can access data, the access level (e.g., embargoed, open, controlled, or closed), and the required data agreements. Because of the sensitive nature of epidemiological data, review by a data access committee and signing of DUAs may be required before access can be granted. This often-lengthy process could be accelerated and facilitated by data systems with online submission, tracking, and approval of requests. Because of their lower sensitivity, metadata should be broadly accessible, even if individual level data are not, and relevant code should be made public through online repositories (e.g., GitHub). This does not necessarily apply to every piece of metadata or code ever generated in a project. As a guiding principle, at minimum datasets and code used to generate findings in scientific publications should be accessible to enable reproducibility.

The preferred method to provide data access is remotely through the internet using standardized authentication and authorization protocols to ensure appropriately controlled access. This is mediated by Hypertext Transfer Protocols (HTTP) over the internet that ensure tight control and digital tracking of what has been accessed, by whom, and when. Remote data access without permanent downloads (i.e., temporary storage of data in memory during analyses or pushing analytics to the data) also facilitates enforcement of DUAs as data use is digitally tracked, avoiding multiple copies of data being distributed across multiple users . Additionally, derived variables that are created as part of new research are created and stored where other collaborators may access and reuse them leading to more consistent variable use across studies and publications. Particularly important is remote access with version control that can track changes to the data and ensure access is provided to the appropriate version (e.g., the most recent version or older versions to replicate previous analyses). All these desirable features are easily available to researchers when data stored in the Cloud are remotely accessed through secure protocols,such as via Application Programming Interfaces (APIs)[19] with authentication and authorization protocols.  Thus, as discussed in more detail later, accessing data in the Cloud greatly facilitates adoption of FAIR practices.

*Interoperable*
Research resources often need to be exchanged, interpreted, and integrated across systems (e.g., combining datasets for pooled analyses). Interoperability refers to the ability of computers to communicate to integrate and use data via automated processes, rather than manually, which is laborious and hard to reproduce (e.g., combining datasets using a computer program rather than manually using spreadsheets). Interoperability is achieved when data from different sources and systems can be integrated, ideally remotely via APIs so that original data sources and versioning can be traced. Interoperability is facilitated when using machine-readable non-proprietary file formats or languages (e.g., CSV, XML, or JSON) and open-source code since these are available to everyone without special licensing. Interoperability should be considered early in the design of epidemiological studies or when developing analytical code, since it is harder to achieve retrospectively.

Data harmonization effortsto combine data from different studies according to common data dictionaries is often time consuming and costly andthe use of data standards, while facilitating integration, restricts the flexibility to generate data in new or preferred ways, potentially limiting innovation[24] .  Commitment to any one data standard during study design still requires mapping and possible remapping in the future to incorporate data from other sources using other standards. Data integration is becoming even more challenging with the increasing volume and variety of data being generated in epidemiological studies. When broadly used by a community, standardized definitions of variables or ontologies can facilitate data integration. However, with some exceptions like the use of ICD10[20] codes for disease classification or ontologies for omics data (e.g., Gene Ontology[21]), their use remains limited in epidemiology. .
. Ontology repositories such as Bioportal[20], or resources to guide the use of standards, databases, and policies such as FairSharing[23], can facilitate these processes.   However, the use of tools such as Schema.org[25] to link between data dictionaries for specific studies or standardized data models better addresses these challenges by facilitating data integration as linked data (described below) and keeping flexibility for data collection[26]. Attaining high levels of interoperability is perhaps the most challenging aspect of FAIR and requires close collaboration among principal investigators, data analysts, and data scientists to bring solutions from computer science to epidemiological research, using standards from the World Wide Web Consortium (W3C) to integrate heterogenous data in the Web[27].

Interoperability of analysis software and workflows[8] is facilitated when code is written using open-source programming languages such as R, Python, Java or JavaScript. The use of open software with public domain licenses such as RStudio promotes re-usability compared to proprietary software such as SAS or STATA. Other features important for code re-usability are described below.

*Re-usable*
The purpose of making data findable, accessible, and interoperable is to enable the optimal re-use of data. Two important impediments for re-usability of data are unnecessary data use restrictions and lack of incentives for data creators. Thus, to maximize re-usability, data should be "as open as possible, as closed as necessary", and access restrictions should be driven by legal and privacy reasons rather than voluntary restrictions.

To ensure the re-usability of the code, analytical pipelines need to be portable, i.e., easy to move from one platform to another. For portability, code dependencies on computational environment (operating system and hardware) and software versions need to be specified in detail, so that the code can be reimplemented and replicated[8,28]. Ideally, the code should also be modular with clear documentation on the operating system as well as versions of installed software packages. There should be detailed specification on the data input and output requirements for the software and workflow, preferably using open standards.

**What are the steps required to adopt increasing degrees of FAIRness in epidemiology?**

A call for adoption of FAIR principles is not an all or nothing proposition. There are varying degrees of epidemiological practice that facilitate re-use of data and code. Even small steps towards this goal could have a big impact [16].  In this section, we explain steps towards achieving increasing "FAIRness" (i.e., adoption of FAIR principles) in two main dimensions that refer to where and how resources are stored. This is illustrated in **Figure 2**, with key steps summarized in **Figure 3**. We use examples of data resources, primarily from cancer epidemiology, to illustrate different degrees of FAIRness.

> A.  *Where are research resources located?*

Moving research resources from on-site to the Web, preferably using Cloud services, progressively facilitates the adoption of FAIR principles as described below.

> ### Step 1: Metadata, data, and code on-site
> Many research resources in epidemiology are stored on-site in private computers or servers. When metadata, data, and code are stored locally, findability and accessibility are low because it requires physical transfers rather than remote access, resulting in data/code silos, multiple copies, and different versions distributed within or across institutions. Interoperability is also low because of difficulties in integrating data sources stored in different local servers, and code dependencies on local computational environments (i.e., having everything you need for code to run properly in different servers). Re-usability of data or code under this scenario tends to be costly and labor-intensive.

Locally stored resources can be made available remotely through Virtual Private Networks (VPN) to increase accessibility. However, VPNs force users to operate in the computational environment of the data hosting institution, which reduces accessibility, interoperability, and ultimately re-usability of data and code. In addition, when working locally, researchers cannot benefit from computing services offered by Cloud providers. Whether remotely accessible or not, using resources stored locally requires institutional IT infrastructure support, security, and maintenance that might not be available in all institutions, particularly those smaller in size or with fewer resources.

***Step 2: Metadata and/or code on Web portals***
Epidemiological studies are increasingly facilitating data sharing through study Web portals, while individual-level data is kept on-site. Portals can provide access to metadata (e.g., study documentation and data dictionaries), data browsers to explore summary-level data, or online submission of data requests. Examples include portals from the NCI Cancer Data Access System ([CDAS](#)) to access data for studies such as the Prostate, Lung, Colorectal and Ovarian Screening Trial ([PLCO](#)). These portals are an important first step in improving findability and accessibility. After approval, datasets can be downloaded through a secure data transfer, but they cannot be remotely accessed through APIs. Providing metadata and code for data harmonization efforts in collaborative projects, such as the NCI Cohort Metadata Repository ([CMR](#)), can also improve interoperability. However, FAIR practices will be limited if datasets remain stored locally, without web-based APIs to allow remote interaction with data.

***Step 3: Metadata and data on Web repositories for data storage and governance***
Depositing individual-level data on trusted Web repositories, along with the necessary metadata to use it, is a big step towards increasing findability since data can be located through Web searches. Cloud-hosted repositories further facilitate accessibility since they typically include centrally managed data access (open or controlled) with monitoring of use. The use of trusted repositories is increasing as funders and scientific journals establish data sharing policies that require depositing individual-level data and code. Even when the data itself requires restricted access, ensuring that metadata can be found through online searches and is broadly accessible is an important step towards FAIRness.

There are many data sharing repositories available to researchers, including general and domain-specific repositories (for a non-exhaustive list, visit [NIH National Library of Medicine](#)). Examples of domain-specific repositories commonly used in epidemiology include the database of Genotypes and Phenotypes ([dbGaP](#)) and the European Genome-Phenome Archive ([EGA](#)) for genomic data. These are controlled-access repositories that have greatly facilitated sharing large-scale genomic data required by policies such as the NIH Genomic Data Sharing effective since 2015. However, availability of other types of epidemiological data in trusted repositories is minimal. This is likely to increase as broader data sharing policies are adopted (e.g., the NIH Policy for Data Management and Sharing in effect in 2023).

Research and government institutions often make public epidemiological datasets available through Web repositories. For example, the NCI Surveillance, Epidemiology and End Results Program (SEER) and the National Health and Nutrition Examination Survey (NHANES) websites allow downloads of non-sensitive data, while restricting the access to sensitive data. SEER currently requires specific software (SEER*Stat), and NHANES requires local access from specific sites (e.g., the National Center for Health Statistics (NCHS) Research Data Center (RDC) workstations) to access sensitive data. Although these are broadly used resources, accessibly and interoperability could be greatly improved by using Cloud-enabled services as described below. Data repositories are evolving to enable FAIR practices by providing data access through secure APIs for remote data analyses without the need for permanent data downloads (e.g., Weekly Provisional Mortality Counts and John Hopkins University COVID-19 Data Repository).

### *Step 4: Cloud-enabled services for data storage, governance, and computation*

Moving metadata, data, and code to the Cloud greatly facilitates FAIR practices because security, governance, and traceability are enhanced by services offered by Cloud providers via APIs. When a research community uses common resources for Cloud data storage and computing to maximize re-usability of resources, it generates what is referred to as a "data commons"[29].  As discussed below, metadata, data, and code do not need to be in the same location or under the same governance, as APIs allow systems to securely communicate to integrate resources remotely.

Commercial Cloud providers offer researchers large capacity for data storage and computing on a pay-as-you-go basis. Cloud services provide as much computational power as needed for fast analyses of large datasets without paying for "idle" time. Incremental cost structures are particularly important for big data analyses where it is costly to move data around for computation, such as in genomics[30].  Remote data access in the Cloud, however, is entirely dependent on having a good internet connection. Mixed systems that span Cloud providers and on-site data centers connected through a VPN, as well as Progressive Web Applications (PWAS) that can work off-line and synchronize when there is internet connection are possible solutions to limited internet access.

From a researcher's viewpoint, commercial Cloud services might increase costs if their institution provides them with high performance computing resources (HPC) at low cost. However, institutional HPC resources are typically not configured to serve data or analytics services to the Web, limiting findability, accessibly, and interoperability. This conundrum is sometimes described as "using the cloud or eventually becoming one". The costs to researchers of using Cloud resources can be mitigated by investments from funding agencies who can negotiate better rates with commercial Cloud providers, such as the US NIH Science and Technology Research Infrastructure for Discovery, Experimentation and Sustainability (STRIDES)[31] Initiative.

Proprietary Cloud enclaves (e.g., DNANexus, Terra and Palantir) are quickly proliferating to provide researchers workspaces for collaborative analyses, as well as infrastructure to share workflows, tools, datasets, and results with other users. Examples of such enclaves are those provided by UK Biobank, All of Us, the National COVID Cohort Collaborative (N3C) and the NHGRI Analysis Visualization and Informatics Lab-space (AnVIL). These proprietary tools facilitate establishing a Cloud-enabled data platform compared to using open-source data ecosystems (e.g., Gen3 or EpiSphere); however, they can be costly to maintain and violate open-source principles for FAIR practices. Another disadvantage is that enclaves often come with governance models that force researchers to stay within the enclave, rather than promoting remote data access from other computational environments. While this could be justified for highly sensitive data with restrictions that do not allow remote access, it can come at a steep price and low interoperability.

B. ***How are metadata and data stored?***

Another critical dimension of FAIRness illustrated in **Figure 2** is how the metadata and data are stored, since this determines how or whether they can be used or re-used by others.

### Step 1: Human-readable formats
Metadata and data stored in file formats that are designed for viewing by humans rather than to be processed by computers are the hardest to use or re-use. Examples include a data table or a questionnaire in a Portable Document Format (PDF) file. Extracting information from a PDF is possible but not trivial.

### Step 2: Machine-readable formats
The use and re-use of data is greatly improved when metadata and data are stored in machine-readable formats that can be automatically read, interpreted, and processed by a computer. Examples include datasets in Microsoft Excel, SAS or Stata formats. These formats can be processed by computers; however, they require proprietary software, limiting accessibly and interoperability.

### Step 3: Non-proprietary formats and languages
Data stored in machine-readable formats that are non-proprietary increase FAIRness because they can be used by anyone with access permission using free software. Traditional tabular non-proprietary data files, such as CSV, have the limitation that they do not contain metadata, which needs to be provided on separate files. Serialization languages such as XML or JSON can embed metadata into the data, addressing this problem. These are basically text files with data and metadata that can be read by humans as well as machines. Most statistical software can read or export data files with metadata in these serialized formats.

***Step 4: Linked data framework***
Linked data is machine-readable data that is interconnected with other data and metadata on the Web. Linked data essentially mimics the way data is organized as interconnected data in the Web of Data[https://pubmed.ncbi.nlm.nih.gov/33479214] using frameworks for data interchange, typically within the Resource Description Framework (RDF)[32]. Compared to a traditional CSV data table, each element in an RDF linked data table is linked to information about the data element using PIDs (i.e., URLs that point to data and metadata on the Web). This is very different from the conventional approach to data annotation where the metadata is kept in separate files in a variety of formats (e.g., data dictionaries in PDF, Word or Excel files). These differences can be explored with the accompanying rdfTable tool and webcast demo. Although linked data is a still a new concept for most epidemiologists, it is increasingly used to better annotate data and cope with the ever-increasing volumes and types of data generated in epidemiology. Note that linked data is not the same as data linkage, which is a method to combine different sources of data for the same unit of observation (e.g., individual, or geographical area).

Relating back to FAIR principles, linked data is machine-readable data and can be found on the Web with all associated metadata (Findable), accessed remotely with version control as open or closed data (Accessible), uses non-proprietary formats/languages that interconnect or integrate data and metadata from different sources remotely without permanent downloads (Interoperable), retaining information about the origin and context the data. Thus, linked data can achieve the highest degrees of FAIRness.

**How can we start making progress towards FAIR practices and systems in epidemiology?**

Modernizing epidemiological data that has been previously collected and stored using traditional approaches can be an expensive and time-consuming endeavor[33]. Therefore, these efforts and their pay offs should be carefully considered, and their value will likely depend on the demand for data sharing.  On the other hand, new efforts should follow FAIR principles from the start. An example is the NCI Connect for Cancer Prevention Study, a new prospective cohort in the USA that is supported by data collection, storage and analysis systems designed following FAIR principles to maximize re-usability of data by the broad scientific community.

Transitioning to using Cloud resources can be daunting for researchers accustomed to local resources. An initial first step that can be made by most researchers is moving data and metadata from local storage to the Cloud using services such as Google Drive, Dropbox, Box or OneDrive/SharePoint. These services are already commonly used to share manuscripts or presentations and are increasingly used to securely store and access sensitive research data. The advantages of these services in support of FAIR practices are summarized in **Box 2**.

Implementation of FAIR principles beyond these initial steps requires collaborations between epidemiologists, data analysts and data scientists with knowledge of computer science. An example of such collaborations is the NCI EpiSphere initiative to develop open-source Cloud-enabled data platforms and tools for adoption of FAIR practices in epidemiology [34], including the following examples:

- Data systems for Connect for Cancer Prevention[35].
- The Mortality Tracker tool [36] for analysis of publicly available data through an in-browser application that provides real time assessment of the effect of COVID-19 on all-cause and cause-specific mortality in the USA[37].
- The PLCO GWAS Explorer tool to exploring and share summary-level GWAS data.
- The Confluence Project Data Platform to facilitate data sharing across consortia through a large multi-ancestry breast cancer GWAS[38].

**Conclusions**

Adopting FAIR practices in epidemiology is needed to address a wide range of challenges faced by epidemiology by improving how data is collected, stored, accessed, and shared. These practices improve daily work and facilitate collaborative analyses and broad data sharing. FAIR practices are becoming a requirement to comply with data and code sharing policies increasingly adopted by many funding agencies and scientific journals to improve the re-usability of data and reproducibility of science[39]. In addition, recruitment and promotion of scientists increasingly involve demonstrating productivity and impact not only through published manuscripts but also through published datasets and code.

Adopting FAIR principles is challenging and requires organizational support, funding, training, and incentives for data originators to share data. But these costs are amply outweighed by the benefits of making research more reproducible, impactful, and equitable by facilitating the re-use of precious research resources by the scientific community.

**Figure 1:** What does FAIR Data and Code Mean in Epidemiological Practice?

## Is my data FAIR ?

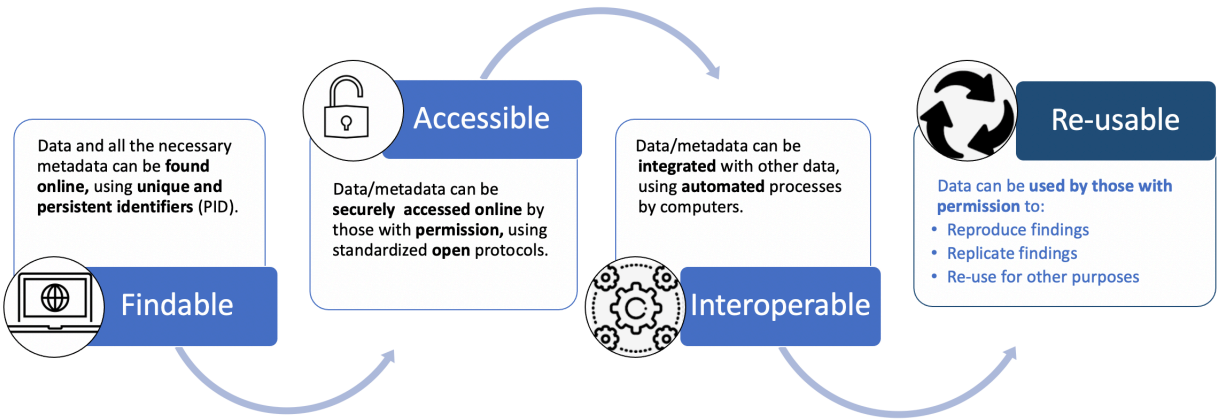

## Is my code FAIR ?

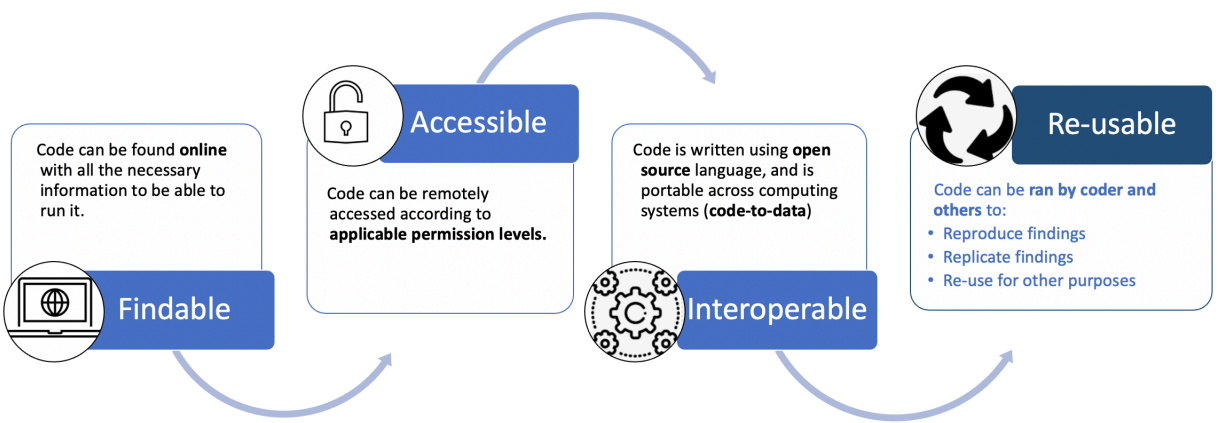

**Figure 2.** Two dimensions of increasing degrees of FAIRness in epidemiology

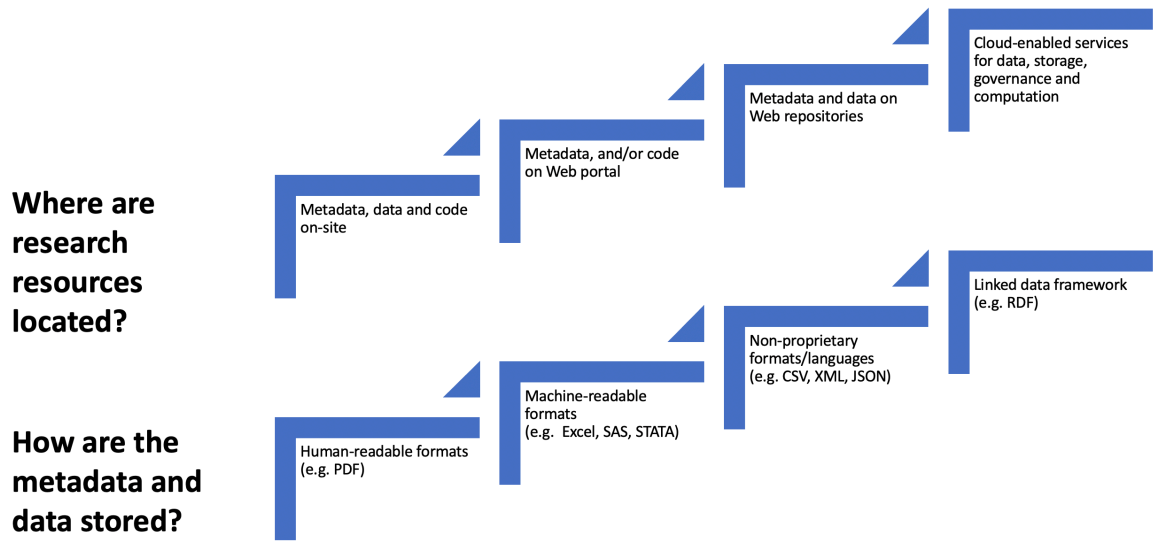

**Figure 3.** Key steps towards increasing degrees of data FAIRness in epidemiology

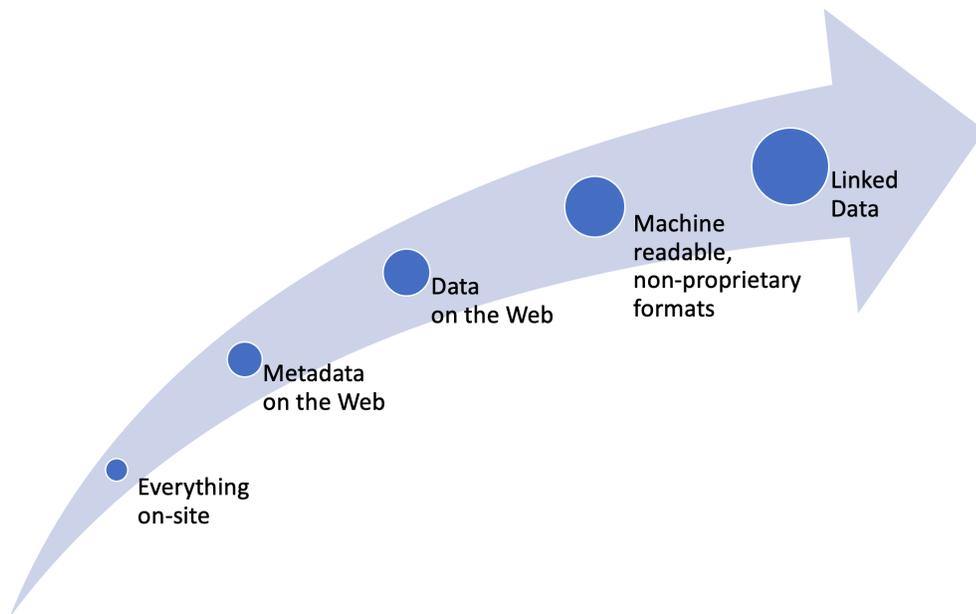

**Box 1. Glossary of Terms**

| Term | Definition |
|---|---|
| Cloud | Servers with software, storage and services accessed over the internet, instead of locally on a computer. Cloud is often used to refer to public Cloud providers (e.g., Google, AWS, Azure) that offer on-demand services and infrastructure shared by many organizations or users through the internet. In contrast, private Clouds are servers with similar characteristic that are only used by one organization. |
| Metadata | Information about the data that enables its use and re-use, including how, when, and by whom it was collected, where to find it, how to adequately use it and how to properly cite it. Examples: study team, study design, data collection instrument (e.g., questionnaire, laboratory assay, EHR), data dictionary, data file information (name, format, size, version, date created/accessed etc.), description of process on how to access the data, data use agreements/licenses, publications using the data, data repository. |
| Machine-readable files | Data or metadata files that can be automatically read, interpreted, and processed by a computer. |
| Data serialization languages | Languages for semi-structured data that facilitate exchange of data across systems (interoperability), by using standardized syntax that contains information about the data. In other words, metadata needed to use data across systems is embedded into the data itself. These languages are both human and machine readable. Examples: XML, JSON |
| Uniform Resource Locator (URL) | Uniform Resource Locator (URL) is a Web address that specifies the location of resources (e.g., data/metadata or code) on the Web, and a mechanism for retrieving it. |
| Globally unique and persistent identifier (PID) | Web address (or URLs) that are long lasting and uniquely identify the location of a resource on the web. |
| Authentication and Authorization protocols | Authentication protocols verify who is the user trying to access the data, and authorization protocols verify that the user has permission to access the data. |
| On-site data or code | Data or code stored locally in personal computers or on servers behind institutional firewalls. |
| Application Programming Interface (API) | A web service that allows applications (e.g., RStudio in your computer) to make a request to a system (e.g., request access to a dataset remotely stored in a server or Cloud) and return a response (e.g. grant access to the dataset). In a nutshell, APIs are an efficient way for computers andmobile devices to communicate. |
| Cloud-enabled data storage and governance services | Consumer-facing computational middle-layer for data storage and governance in the Cloud. Examples: Google Drive, Dropbox, Box, OneDrive/SharePoint |

| Data commons | Shared resources for Cloud data storage and computing designed for re-usability by a research community |
|---|---|
| Data egress costs | Fees charged by Cloud providers when data stored under their services is exported. |
| | |

**Box 2. Advantages of remote data access using Cloud-enabled services (e.g., Google Drive, Dropbox, Box or OneDrive/SharePoint)**

- Findable:
  - Data and metadata files get automatically assigned a unique URL (PID) when uploaded to a Cloud-enabled service.
  - Online search functions to find files from any location by those with permissions.
- Accessible:
  - Data can be securely accessed remotely via APIs to run analyses in the analyst computer or server, without permanent data downloads.
  - Management of access permissions is user friendly, providing full control of who can access the data.
  - File usage statistics track data usage by owner and others with access, providing a historical record of all modifications to the data as researchers work with it.
  - Version control keeps track of data modifications by owner and others with access. This avoids creating multiple files with different versions of the data.
- Interoperable:
  - Datasets from different owners stored in different accounts (e.g., two datasets in folders owned and managed by two different investigators) can be integrated using APIs without moving the data.
  - APIs can be used to read a subset of data in a large dataset needed for a given analysis.
- Re-usable:
  - Data can be securely shared for re-use by others through access permissions
  - Data stays under the governance of the owner with clear data provenance so that owners can get appropriate recognition